\documentclass{amsart}
\usepackage{graphicx}
\newtheorem{thm}{Theorem}[section]

\theoremstyle{definition}
\newtheorem{defn}[thm]{Definition}
\theoremstyle{remark}

\numberwithin{equation}{section}

\newcommand{\abs}[1]{\left\vert#1\right\vert}

\begin{document}

\title[]{Phase Spaces in Special Relativity: Towards Eliminating Gravitational Singularities}
\author{Peter Danenhower}
\address{Langara College\\
Vancouver B.C.\\
Canada, V5Y 2Z6} \email{pdanenho@langara.bc.ca}

\thanks{I wish to thank Cisco Gooding for kindly reading over several drafts
of this paper and offering valuable comments and insights that
clarified my understanding in some sections of this paper.}

\dedicatory{}
\begin{abstract}
This paper shows one way to construct phase spaces in special
relativity by expanding Minkowski Space.  These spaces appear to
indicate that we can dispense with gravitational singularities. The
key mathematical ideas in the present approach are to include a
complex phase factor, such as, $e^{\i\phi}$ in the Lorentz
transformation and to use both the proper time and the proper mass
as parameters. To develop the most general case, a complex parameter
$\sigma=s+im$, is introduced, where $s$ is the proper time, and $m$
is the proper mass, and $\sigma$ and $\frac{\sigma}{|\sigma|}$ are
used to parameterize the position of a particle (or reference frame)
in space-time-matter phase space.  A new reference variable,
$u=\frac{m}{r}$, is needed (in addition to velocity), and assumed to
be bounded by 0 and $\frac{c^{2}}{G}=1$, in geometrized units.
Several results are derived: The equation $E=mc^2$ apparently needs
to be modified to $E^2=\frac{s^2c^{10}}{G^2}+m^2c^4$, but a simpler
(invariant) parameter is the ``energy to length" ratio, which is
$\frac{c^4}{G}$ for any spherical region of space-time-matter. The
generalized ``momentum vector" becomes completely ``masslike" for
$u\approx 0.79$, which we think indicates the existence of a maximal
gravity field.  Thus, gravitational singularities do not occur.
Instead, as $u\rightarrow 1$ matter is apparently simply crushed
into free space. In the last section of this paper we attempt some
further generalizations of the phase space ideas developed in this
paper.
\end{abstract}
\maketitle
\section{Introduction}
Phase spaces are a common and important way to model physical
systems.  For example, a harmonic oscillator has the energy equation
$E=\frac{p^2}{2m}+\frac{kx^2}{2}$, which expresses the energy of the
oscillator (pendulum, mass on a string, etc) as a function of the
position and momentum. More generally phase spaces, usually called
symplectic structures or symplectic spaces, have been extensively
studied.  The most basic symplectic structure is a smooth Manifold
with a closed, non-degenerate 2-form, such as, $\omega=dp_i\wedge
dq^i$.  Equivalently, given a manifold of all possible
configurations of a system, the phase space for the system is the
cotangent bundle. See ~\cite{dMdS98} for an introduction to
symplectic structures.

Phase spaces within General Relativity have also been studied, for
examples see ~\cite{aH80,AT02}.  Unfortunately, while interesting
this work does not seem to lead to any resolution of important open
questions in general relativity, such as cosmic censorship or the
validity of gravitational singularities.  Even expanding the phase
spaces to complex number phase spaces does not seem to be sufficient
to answer these questions, but for an example of the complex
treatment, see ~\cite{BFP80}.

Although, general relativity and especially special relativity, have
survived intense experimental and observational testing
~\cite{MTW73, rW84, sW72, cW98, gEF99, BS00, CMS99,tD00}, the
question of the existence of gravitational singularities is still
unresolved.  My sense is that most astro-physicists and general
relativist would prefer that gravitational singularities were not
predicted by general relativity, but are willing to accept them if
necessary, because general relativity works so well otherwise. Thus,
we have assumed in this paper that attempting to eliminate the
singularities predicted by general relativity is a desirable result
(see for example, ~\cite{MTW73}, chapter 44).

The present paper is motivated by the desire to try to eliminate
gravitational singularities (without changing anything else in a
measurable way). It is somewhat surprising that we seem to be able
to make progress within the framework of a complex phase space set
in special relativity, especially as the present construction is not
intended to incorporate gravity. The main difference between the
present work and previous efforts to build relativistic phase spaces
is the inclusion of the proper mass as a dynamic variable, as
opposed to a constant scale factor.  The consequences of using the
proper mass as an additional parameter (along with proper time) seem
to be significant enough to warrant beginning this study within the
frame work of special relativity.

The present work begins with Minkowski space and changes this to a
complex phase space, a space-time-matter configuration space, by
introducing a complex ``phase" factor into the Lorentz
transformation. The special relativistic ``four" momentum and
``four" position are combined into a single vector (1-form) in
$C^4$. Of course, the proper mass is constant, so we have to work
out the transformation equations (complex Lorentz transformation)
for an arbitrary mass, then substitute the actual mass of interest.
Actually, we can show that there is a canonical value of the proper
time as well, so that proper time and proper mass are not that
different after all.

To make the phase factor dynamically significant requires another
parameter to play a role similar to the velocity. This is a another
key difference between the present work and previous phase space
developments in general relativity.  The parameter needs to have
dimensions of mass over length (or coordinate time). Fortunately,
this ratio is already a parameter of some interest in general
relativity (although only for extremely massive objects), so we have
assumed the mass to length ratio of general relativity to be the
parameter we need.  Then gravitational effects become significant in
an extreme relativistic limit.

We have tried to introduce the phase factor into the Lorentz
transformation in the most general possible way.  However, it turns
out to be very simple, i.e., of the form $e^{i\phi}$. Nevertheless,
we have proceeded with the general derivation given in this paper,
because some of the results of the derivation are used in the
further developments in the last section. The outline of this paper
is as follows:
\begin{enumerate}
    \item  State the postulates of the Phase Space.
    \item Decide what forms need to be invariant and derive the
    conditions on the complex Lorentz group.
    \item Derive the Phase Space Lorentz transformation.
    \item Consider some of the consequences of these equations.
    \item Consider some generalizations.
\end{enumerate}

\section{Postulates of the Special Relativistic Phase Space}
Since the phase space constructed in this paper is built within
special relativity, we certainly need the postulates of special
relativity.  Briefly, we can recall these to be:   1. The laws of
physics should have the same form (in Cartesian coordinates) for all
Lorentz observers, and 2. The speed of light $c$, is constant for
all Lorentz observers.

We need another assumption to make further progress, namely the
gravity constant, $G$ is also invariant for all Lorentz observers. I
assume that most readers will find this to be a reasonable
assumption that is pretty much already assumed in general
relativity.  One can imagine that different Lorentz observers could
verify this assumption by doing extremely weak field experiments, so
that the Lorentz character of the reference frames was not
disturbed.

In addition, for the following calculations we have used geometric
units chosen so that $c=1$ and $G=1$.  We have denoted the mass to
length ratio with the parameter, $u=\frac{m}{r}$.  Then, in
geometrized units, $0\leq u \leq 1$. The transformation equations
derived below require this restriction on $u$, similar to the
restrictions on $v$. The particular choice, $\frac{c^2}{G}=1$ is
made to normalize $u$ as simply as possible. This is not quite the
Schwarzschild radius predicted by general relativity, which is
$\frac{c^2}{2G}=1$. Whether or not there is a two in denominator (or
some other scalar factor is a matter to be decided by experiment, so
for the purposes of this paper I have omitted the two for
simplicity.

I am well aware that the assumption that $0\leq u \leq 1$ is likely
to be disputed or rejected by some readers, but I think this
requirement is simply a reflection of the cosmic censorship
conjecture: what happens inside the Schwarzschild radius isn't
observable from the outside, so $u$ will be observed to be bounded
above by one (the development given here does not forbid $u=1$). In
any case, there is no question that for ordinary objects $u$ is
extremely small, mostly not even measurable. For example, for the
earth, $u\approx10^{-10}$, and for a 1 kg ball of radius 1 m,
$u\approx10^{-27}$.  $u$ is still smaller for elementary particles.

I have found little discussion in the literature of the parameter,
$u$.  The internet motion mountain physics text does discuss $u$,
but treats this parameter as force.  I don't think this is a good
idea, because force is such a difficult concept in special
relativity.

\section{Invariant Forms}
To develop the phase space idea in this paper, we have assumed that
the volume form, $dV=dy^0\wedge dy^1\wedge dy^2$ is invariant with
respect to the allowed coordinate transformations. For the time
being we are considering a 3 dimensional space consisting of two
space dimensions and one time dimension, with the usual index
conventions (0 for time, 1 and 2 for space dimensions). To keep the
problem as simple as possible the direction of relative motion is
assumed to be along one of the space coordinates.  The second space
dimension may not be necessary, but we have included it for reasons
that will become clear as the derivation proceeds.

We begin with the ``position" vector, $\mathbf{y}$, and the two
tangent vectors, $\mathbf{y}_{,s}$ and $\mathbf{y}_{,m}$. Here
``position" vector means position in the complex phase space, not
the physical position.  In these expressions, $s$ is the proper
time, and $m$ is the proper mass of the particle. In addition, I
have assumed that we can extract the magnitude of the two form
representing the cross product of tangent vectors (roughly the
symplectic 2-form discussed above), using Hodge star duality. Since
we don't have an explicit metric defined, the definition of the
Hodge star dual is quite significant, but I have defined this in the
natural way, i.e., so that we get the answer we would expect if we
had a metric. Apparently the definition that is used below induces a
metric on the space, but the significance of this isn't entirely
clear to me.

In any case, we have assumed the surface area form is invariant with
respect to the allowed coordinate transformations. The following
derivation is probably much more complicated than it needs to be.  I
have done the derivation this way to set the stage for generalized
equations of motion. However, to obtain just the complex Lorentz
transformations in a simple way, we can just multiply the usual
components of the Lorentz transformation by a complex ``phase"
factor of modulus 1, and assume a complex conjugated, Lorentz
metric.

Continuing with the derivation, from the tangent vectors, we can
construct the area 2-form (symplectic 2-form),
\begin{equation}
dA=\delta_{kn}^{01} y^k_{, s}y^n_{, m} dy^0dy^1+\delta_{kn}^{02}
y^k_{, s}y^n_{, m} dy^0dy^2+ \delta_{kn}^{12} y^k_{, s}y^n_{, m}
dy^1dy^2
 \end{equation}

 In this expression I have used the generalized Kronecker delta
 (anti-symmetric in $k$ and $n$, so $\delta ^{01}_{01}=1, \delta ^{01}_{10}=-1$
 and $\delta_{kn}^{01}=0$ for any other choice of $k$ and $n$) and there is,
 of course, no sum
 on the fixed numerical indices.  I have omitted the wedge product symbol for brevity.
 For the rest of this section I have assumed the wedge product unless otherwise stated.

 Next we need to define the Hodge
 star duals of the basis 2-forms, so we can find $\ast dA$.\\

\begin{defn}
The Hodge star duals of the basis two forms are defined as follows:
$\ast (dy^1\wedge dy^2)=-dy^0$, $\ast (dy^2\wedge dy^0)=dy^1$, $\ast
(dy^0\wedge dy^1)=dy^2$
\end{defn}

Notice that I have defined $\ast (dy^1\wedge dy^2)$ to have the
opposite sign of what might be expected.  This is to avoid having
to use $i$ to keep the contributions to the volume formally
positive. In addition, to this definition, I have also assumed
throughout this paper that to find the Hodge star dual of a 1 or 2
form we have to use the conjugated transpose form of the
components (this will be clear below). Thus, we have the following
expression for $\ast dA$:
\begin{equation}
\ast dA=\delta_{kn}^{01} (y^k_{, s}y^n_{, m})^{\dagger}
dy^2+\delta_{kn}^{20} (y^k_{, s}y^n_{, m})^{\dagger} dy^1-
\delta_{kn}^{12} (y^k_{, s}y^n_{, m})^{\dagger} dy^0
 \end{equation}

In this expression the $\dagger$s indicate complex conjugated
transpose. Finally, we can extract the magnitude of the ``symplectic
2-form" as the component of $\ast dA\wedge dA$.  I have called this
$L^2$ for reasons that will be explained later.

\begin{equation} L^2=\delta_{kl}^{01}\delta_{np}^{01}
(y^k_{, s}y^l_{, m})^{\dagger}y^n_{, s}y^p_{, m} +\delta_{kl}^{20}\delta_{np}^{20}
(y^k_{, s}y^l_{, m})^{\dagger}y^n_{, s}y^p_{, m}- \delta_{kl}^{12}\delta_{np}^{12}
(y^k_{, s}y^l_{, m})^{\dagger}y^n_{, s}y^p_{, m}
 \end{equation}

 To derive the complex Lorentz transformation we want to find the sub-group of
 $SL(3)$ (special linear group in 3 dimensions) that keeps the righthand side of
 equation 3.3 invariant.  We start with the transformation equations,
 $y^k=B^k_j\hat{y}^j$, where $B^k_j$ is a $3\times 3$ matrix (with constant complex entries),
 such that $\abs{\mbox{det}(B^k_j)}=1$.  Substituting the transformation
 equation into the righthand side of equation 3.3 and subtracting the righthand side
 of equation 3.3, we get:
\begin{equation}
(\delta_{kl}^{01}\delta_{np}^{01}-\delta_{kl}^{12}\delta_{np}^{12}+
\delta_{kl}^{20}\delta_{np}^{20}) \times
[(B^k_{\alpha}\hat{y}^{\alpha}_{,s}B^l_{\beta}\hat{y}^{\beta}_{,m})^{\dagger}
B^n_{\rho}\hat{y}^{\rho}_{,s}B^p_{\sigma}\hat{y}^{\sigma}_{,m}-
(\hat{y}^k_{,s}\hat{y}^l_{,m})^{\dagger}\hat{y}^n_{,s}\hat{y}^p_{,m}]=0
\end{equation}

In equation 3.4, both Greek and Latin indices are summed over 0,1,2.  I am using
Greek indices simply because Latin indices are in short supply and also for editing
purposes.  The multiplication operations are all just complex numbers, so we can
rearrange equation 3.4 to get:
\begin{equation}
(\delta_{kl}^{01}\delta_{np}^{01}-\delta_{kl}^{12}\delta_{np}^{12}+
\delta_{kl}^{20}\delta_{np}^{20}) \times
[(B^k_{\alpha}B^l_{\beta})^{\dagger}B^n_{\rho}B^p_{\sigma}(\hat{y}^{\alpha}_{,s}\hat{y}^{\beta}_{,m})^{\dagger}
(\hat{y}^{\rho}_{,s}\hat{y}^{\sigma}_{,m})-
(\hat{y}^k_{,s}\hat{y}^l_{,m})^{\dagger}\hat{y}^n_{,s}\hat{y}^p_{,m}]=0
\end{equation}

In this equation the $\dagger$s indicate the entries in the conjugated transpose,
$B^{\dagger}$.  So for example, $(B^k_l)^{\dagger}$ is the entry in the $k^{th}$ row
and $l^{th}$ column of $B^{\dagger}$.  There is, of course, no matrix multiplication
in this equation, so we can freely move the transpose operation inside the brackets
if we wish.

We seek the minimal conditions on $B^k_l$ that will satisfy this equation without
restricting the values of $y^k_{,s}$ or $y^k_{,m}$. By making a fixed choice of
$k,l,n$ and $p$ in the last term (in the square brackets) and then making the same
choice for $\alpha , \beta , \rho$ and $\sigma$, we can then sum over $k,l,n$ and
$p$ in the first term to get the following equations that constrain $B^k_l$.

\begin{gather}
\triangle (B^2_2)^\dagger\triangle B^2_2+\triangle (B^1_2)^\dagger\triangle B^1_2
-\triangle (B^0_2)^\dagger\triangle B^0_2-1=0,\\
\triangle (B^2_1)^\dagger\triangle B^2_1+\triangle (B^1_1)^\dagger\triangle B^1_1
-\triangle (B^0_1)^\dagger\triangle B^0_1-1=0,\\
\triangle (B^2_0)^\dagger\triangle B^2_0+\triangle (B^1_0)^\dagger\triangle B^1_0
-\triangle (B^0_0)^\dagger\triangle B^0_0+1=0.
\end{gather}

In these equations, $\triangle B^k_l$ indicates the cofactor of the entry in B in
the $k^{th}$ row and $l^{th}$ column.  Of course, we also require that
$detB=e^{i\theta}$, for some phase angle, $\theta$.

If we set $B=\left[
\begin{smallmatrix}
b_{00} & b_{01} & 0\\
b_{10} & b_{11} & 0\\
0 & 0 & 1
\end{smallmatrix}\right]$,
then we can substitute into equations 3.6 - 3.8, and $detB=e^{i\theta}$, to get:

\begin{gather}
\overline{(b_{00}b_{11}-b_{01}b_{10})\!}\,(b_{00}b_{11}-b_{01}b_{10})-1=0,\\
-\overline{b_{01}\!\!}\;b_{10}+\overline{b_{00}\!\!}\;b_{00}-1=0,\\
-\overline{b_{11}\!\!}\;b_{11}+\overline{b_{10}\!\!}\;b_{01}+1=0,\\
b_{00}b_{11}-b_{01}b_{10}=e^{i\theta}.
\end{gather}

The choice of $b_{22}=1$ and other entries involving an index of 2
equal to zero is certainly a special choice (for example, certainly
$b_{22}=e^{i\phi}$ would be more general), but I am calculating
essentially a two dimensional case. Notice that equations 3.9 and
3.12 are essentially the same statement: $\abs{detB}=1$. Our next
task is to try to make the matrix B physically reasonable by
deciding (guessing) how the usual Lorentz transformation appears in
the matrix B. The simplest is to attach a complex number factor to
each entry in the usual Lorentz transformation:
\begin{equation}
B=\begin{bmatrix} (e+if)\gamma & -v\gamma (c+id) & 0\\
-v\gamma (a+ib) & \gamma (a+ib) & 0\\
0 & 0 & 1
\end{bmatrix}\end{equation}
Substituting, this B into equations 3.9 - 3.12,we get the following equations:

\begin{gather}
-(a^2+b^2)\gamma ^2+v^2\gamma ^2(a-ib)(c+id)+1=0\\
-v^2\gamma ^2(c-id)(a+ib)+(e^2+f^2)\gamma ^2-1=0\\
(a^2+b^2)\gamma ^2[(e+if)-v^2(c+id)][(e-if)-v^2(c-id)]-1=0
\end{gather}

A straight forward calculation yields,
\begin{equation} e=a,\: f=b,\: c=\frac{((a^2+b^2)\gamma^2-1)a}{v^2\gamma
^2(a^2+b^2)},\: \mbox{and}  \: d=\frac{((a^2+b^2)\gamma^2-1)b}{v^2\gamma
^2(a^2+b^2)}.
\end{equation}

We are left to decide how to eliminate the last variable.  Substituting equations
3.17, using the matrix 3.13, into 3.12 gives the basic conditions on $a$ and $b$,
which simplifies to:
\begin{equation}
\frac{(a+ib)^2}{a^2+b^2}=e^{i\theta}
\end{equation}
Unfortunately, this equation doesn't seem to say much.  It is
tempting to suppose that $a^2+b^2=1$, because this choice makes B
symmetric, since in that case $c=a$ and $d=b$, but it is also
possible to find a stronger argument:  The magnitude of the tangent
vectors, $y^k_{,s}$ and $y^k_{,m}$ (using the Hodge star duality in
definition one) should not be explicitly dependent on $v$.  Thus,
$\abs{y^k_{,s}}^2=-(a^2+b^2)\gamma ^2+\frac{((a^2+b^2)\gamma
^2-1)^2}{v^2\gamma ^2(a^2+b^2)}$, and a short calculation (using
$\gamma ^2=\frac{1}{1-v^2}$, of course) shows that the only solution
to $\frac{\partial}{\partial v}(\abs{y^k_{,s}}^2)=0$ is
$(a^2+b^2)=1$.  Thus, the whole exercise boils down to multiplying
the usual Lorentz transformation by a complex phase factor,
$\exp{(i\phi)}$.

In any case, to finish the calculation and interpret $\exp{(i\phi)}$
in terms of $u$, the next decision we have to make is whether to
solve for $a$ or for $b$.  If we let $u=\frac{m}{r}$ be $a$ or $b$,
as the case may be, one way we get $(a+ib)=u+i\sqrt{1-u^2}$ and the
other we get $(a+ib)=\sqrt{1-u^2}+iu$.  In the first case the usual
(non-phase space) relativistic limit is recovered when $u=1$ (since
the imaginary part of the factor, $u+i\sqrt{1-u^2}$ needs to be zero
in this limit). In the second case, the usual relativistic limit
occurs when $u=0$. Of course, there is really no reason why the
relativistic limit could not occur when both factors are pure
imaginary, so that these considerations would be reversed. Thus, the
simplest choice (I think it does not matter, in fact) is to let
$(a+ib)=\sqrt{1-u^2}+iu$.

Thus, to summarize our final result:
\begin{equation}
B=\begin{bmatrix} \gamma (u+i\sqrt{1-u^2}) & -v\gamma (u+i\sqrt{1-u^2}) & 0\\
-v\gamma (u+i\sqrt{1-u^2}) & \gamma (u+i\sqrt{1-u^2}) & 0\\
0 & 0 & 1
\end{bmatrix}\end{equation}

\section{Discussion and Consequences of the Phase Space Lorentz Transformation}
\subsection{Effective Metric}

At this point it is worth taking some time to try to interpret the
transformation physically.  This isn't easy, since the complex phase
space we have constructed appears to be much more of a configuration
space than even the usual Minkowski space.  In addition, the
corrections appear to be virtually unmeasurable excepting, in the
relativistic phase space limit, which entails enormous gravity
fields. Apart from this environment not normally being considered
within the context of special relativity, the high gravity
environment is so beyond our ordinary experience that it is
difficult to be sure exactly what the equations are telling us.

In the first place it is useful to ask if we effectively have a
metric. Using the Hodge star dual in definition 3.1, the matrix B in
equation 3.19 is clearly Lorentz orthogonal.  So apparently
definition 1 together with the condition, $\frac{\partial}{\partial
v}(\abs{y^k_{,s}}^2)=0$, are sufficient conditions to induce a
complex conjugated metric. So for the rest of the paper we will
mostly just use the induced metric to simplify the considerations.
\subsection{Complex Phase Space, Energy Equation and Space-Time-Matter}

Following the usual procedure in special relativity, we can define a
``four-velocity" (actually only a ``two-velocity" here), by
\begin{defn}
\begin{equation}
\mathbf{U}=(\sqrt{1-u^2}+iu)\gamma<1,v>
\end{equation}
\end{defn}

It is not immediately obvious how to construct the phase space
``four momentum" from the four velocity.  It is tempting to set
$\mathbf{P}=\mathbf{U}(s+im)$, and in fact, we set several drafts of
this paper with this definition.  However, this definition will not
do for reasons that will be explained later.  Instead we need to use
a scaled parameter, $\frac{s+im}{\sqrt{s^2+m^2}}$ and construct the
``momentum per unit length" vector:

\begin{defn}
\begin{eqnarray}
\hat{\mathbf{P}}=<y^0,y^1>=(\sqrt{1-u^2}+iu)\gamma<1,v>\sigma \\
=(\sqrt{1-u^2}+iu)\gamma<1,v>\frac{s+im}{\sqrt{s^2+m^2}} \nonumber
\end{eqnarray}
\end{defn}

Recall that in this expression $s$ is the proper time and $m$ is the
proper mass of the particle or object under consideration.
$\gamma<1,v>$ is the usual ``2 dimensional" tangent vector.  For the
rest of the paper I have adopted the convention of designating
``expressions per unit length" by hatting the usual notation.
``Expression" means the momentum, energy, etc.  In addition, we have
not concerned ourselves too much with the distinction between
tangent vectors  (four velocity) and cotangent vectors (four
momentum), since in special relativity the distinction is not too
important.

Clearly, with this definition $\hat{\mathbf{P}}$ is the position
vector in a complex phase space. This is a central aspect of the
current phase space construction: an important feature of the usual
theory of special relativity is the merging of space and time. For
example, the separation between events has space and time
components, which are different for different observers. In the
present development we are trying to accomplish a further merging,
namely, of mass and space-time to form space-time-matter.

There is some precedent for doing this already from general
relativity and general relativistic phase space, but I don't think
the merging in general relativity is complete.  We represent gravity
with curvature, but in my opinion the curved space is still a
representation of the physical space (with masses and forces - we
can talk about curved space-time around the Sun, but the Sun is
obviously still there), and not an actual merging of space-time and
matter. The Einstein field equations serve as a ``code" for
converting matter and energy into a curved space-time model, and
back again, but there isn't an actual merging.

In any case, it is of some interest to multiply-out the two complex
factors in equation 4.2 to get:
\begin{equation}
\hat{\mathbf{P}}=\frac{\left[(s\sqrt{1-u^2}-um)+i(m\sqrt{1-u^2}+us)\right]}{\sqrt{s^2+m^2}}\gamma<1,v>
\end{equation}

In this expression, recall that for ordinary terrestrial physics or
fundamental particle physics, $u$ and especially $u^2$ are
vanishingly small, so that to great accuracy for terrestrial
experiments,
$\hat{\mathbf{P}}=\frac{s+im}{\sqrt{s^2+m^2}}\gamma<1,v>$. In
addition, in this limiting case, $\sqrt{s^2+m^2}\approx s$, so we
can recover the usual momentum vector as
$\mathbf{P}=(s+im)\gamma<1,v>$.  In this limit the space-time and
mass aspects of $\mathbf{P}$ have been separated into the usual
position vector and momentum vectors of special relativity. So in
this limit we have only accomplished the merging of space-time and
mass transformation equations into a single expression, essentially
a one parameter special relativistic phase space.  This phase space
is the special relativistic analogue of classical phase space.

Before discussing this further it is worth calculating the energy
per unit length of the particle.  Using the special relativistic
equation, $E^2=-\mathbf{P}\cdot\mathbf{P}$, we have immediately,

\begin{equation}
E^2=s^2+m^2
\end{equation}

Recall that we are using a complex conjugated, Lorentz metric with
$\eta_{00}=-1$, so that the ``four velocity" has constant magnitude
equal to -1.  It is interesting to insert the clusters of constants
needed to make this equation have metric units (kilograms, metres
and seconds).  In this case,
\begin{equation}
E^2=\frac{s^2c^{10}}{G^2}+m^2c^4
\end{equation}

The second term in equation 4.5 is just the usual relativistic mass
energy term, but the first term is new.  Apparently this means that
as time passes, we must include the energy of free space inside a
spherical ball of radius $r=sc$, i.e., keep $r$ on the null cone. If
we ignore the mass energy term (which is minuscule by comparison for
terrestrial experiments), then we have $E=\frac{sc^5}{G}$, so the
energy to radius ratio of a ball of free space with radius $r=sc$
has the colossal value of,
\begin{equation}
\hat{E}=\frac{c^4}{G}\approx1.21\times10^{44}\mbox{joules per metre}
\end{equation}

Thus, in the present phase space construction, free space has an
energy to radius ratio equal to that of matter that has been
compressed to the limiting ratio of $\frac{c^2}{G}$. The obvious
conjecture is that in this limit matter has been crushed into empty
space.  I don't think this is so far fetched, even from the point of
view of general relativity, since in the latter theory matter is
predicted to be crushed into a space-time singularity.  The current
approach dispenses with the need for a singularity, by assumption,
to keep $u$ bounded. The concept of energy of space-time gives the
mass energy some where to go, so that we do not need a singularity.
Nevertheless, there are clearly problems with energy conservation,
discussed presently.

At this point we can understand what is wrong with using $E$ instead
of $\hat{E}$.  Equation 4.5 seems to say that the energy of a region
of space-time-matter, with a value of $u\approx 1$, has a total
energy of $E=\frac{\sqrt{2}sc^5}{G}$, so that
$\hat{E}=\frac{\sqrt{2}c^4}{G}$.  This won't do, because equation
4.6 says that $\hat{E}$ for free space does not have the factor of
$\sqrt{2}$.  Thus, the conclusion I have drawn from this is simply
that energy is not a relativistic phase space invariant, and instead
the important (invariant) quantity is $\hat{E}$.  If we compute
$\hat{E}$ directly using $\hat{\mathbf{P}}$, we get $\hat{E}=1$, as
expected. This says the total energy per unit length (of radius) of
any ball of space-time-matter is $\frac{c^4}{G}$.  Evidently energy
is not conserved in this relativistic limiting case.  We will
returned to this issue.

A less obvious consequence of equation 4.6, and no doubt much more
controversial, is that the external gravity field of a ``black hole"
is apparently zero.  This follows from the direct experiential fact
that the gravity field of free space is zero.  Since matter at the
extreme limit of $\frac{c^2}{G}$ appears to have been crushed into
free space, there is no external gravity field.  The immediate
objection is that there appears to be ample evidence that  extremely
strong gravity fields (too strong to be neutron stars) are common in
the universe. However, below we have shown that equation 4.3 seems
to suggests that there is a maximal gravity field with $u$ several
times larger than a neutron star, but still well below the critical
value of $u=1$.

\subsection{Addition formula for $u$}

At this point, it would be helpful to try to understand the
parameter $u$ a little better.  Apparently, $u$ is quite different
from $v$, since each observer can measure $u$ (for their own
reference frame or another observer) within their own frame, i.e.,
$u$ appears to be absolute, whereas, $v$ is relative between
observers.  Nevertheless, great caution is needed, since we have so
little direct experimental data that illuminates the property being
measured by $u$.  In any case, since there is an upper bound for
$u$, we do have to assume there is a ``relativistic addition" rule.
We can calculate the exact expression for addition of $u$ (using a
combination of the polar and algebraic forms for the complex factor
of modulus one):
\begin{eqnarray}
\mbox{exp}(i\phi)=\mbox{exp}i(\theta_1+\theta_2)=
\mbox{exp}(i\theta_1)\times\mbox{exp}(i\theta_2)= \nonumber \\
(\sqrt{1-u^2_1}+iu_1)(\sqrt{1-u^2_2}+iu_2)= \nonumber \\
\sqrt{1-u^2_1}\sqrt{1-u^2_2}-u_1u_2+i(u_2\sqrt{1-u^2_1}+u_1\sqrt{1-u^2_2})
\label{line2}
\end{eqnarray}

If we compare the real and imaginary parts of this last expression
with the form, $\sqrt{1-w^2}+iw$, it is straight forward to verify
that
\begin{equation}
w=u_2\sqrt{1-u^2_1}+u_1\sqrt{1-u^2_2}
\end{equation}

satisfies both the real and imaginary parts.  It can also be readily
verified that $0\leq w\leq 1$.  Thus, equation 4.8 is the phase
space addition formula for $u$.  Notice that for terrestrial values
of $u_1$ and $u_2$ (minuscule), $w\approx u_1+u_2$ to a high degree
of accuracy.  We need to be careful how we interpret this equation:
apparently this equation would apply to say an observer at the
center of the earth who made a further measurement of an object at
the surface of the earth. Then $u_1$ would be for the earth and
$u_2$ would be for the object. In particular, we would apparently
not use equation 4.8 to measure the results of say, a collision of
two neutron stars (assuming the stars coalesced into one star).

The phase space development gives us a way to make sense of $u$ for
a small irregularly shaped object for which the value of $r$ to be
used in computing the ratio $\frac{m}{r}$ is apparently unclear. The
value of $r$ always makes sense in the phase space approach as
$r=sc$, where $s$ is the minimum possible proper time needed to
measure $u$. So $r$ is on the light cone.  Incidently, this
definition makes $s$ have an inherent value for each object, in the
same way that $m$ apparently seems to have an inherent value. So for
example, a steel rod of length $L$ would be understood to have
$r=\frac{L}{2}$ (observed from the center in the rest frame, which
makes $s$ as small as possible), because according to the phase
space development, the rod consists of its mass and shape as a rod,
but also includes a ball of space-time with proper time
$s=\frac{L}{2c}$.
\subsection{Possibility of a Maximal Gravity Field}

Before discussing the possibility of a maximal gravity field, I
would like to suggest that the result of general relativity that
assumes that a black hole has an enormous gravity field depends on a
principle that I think is not relativistic:  the usual explanation
is that the space-time around the black hole is permanently curved
as the matter collapses. A principle such as this is needed because
the source of the gravity field inside the event horizon has no way
of communicating with the external field (at least within the
context of general relativity alone and assuming either weak or
strong cosmic censorship). In my opinion this is not a general
relativistic principle because it is not local: according to the
usual explanation the space around the event horizon stays curved
for all time, i.e. for an extended separation of space-time (at
least in absence of encounters with other gravity fields or quantum
effects). I think the explanation that is true to relativity is that
space-time must constantly be informed from the (local) source of
the gravity field. Since the source is not available (it is inside
the event horizon), the field vanishes.

I am well aware that gravitons, assumed to be the ``messenger" for
gravity, are not predicted to radiate unless the field changes.  I
think this just means we will have to devise a more sophisticated
model of how gravity fields are sustained, and electrostatic fields
for that matter.  The issue here also raises the old debate of
``gravity is curvature of space time" versus ``curvature of space
time is a representation of gravity". The present approach is
consistent only with the representation point of view, because the
physics is gone behind the event horizon, so there should be no
curvature as well. I believe the ``gravity is curvature" view is too
extreme for the simple reason that we can still go into the lab and
do all sorts of non-gravity physics, say measure a current or a
temperature, so it would be very peculiar if gravity was not also
physical.  Thus, general relativity is surely a configuration space,
a way to represent gravity geometrically:  we can do either physics
or geometry.  Thus, if the physics is gone (behind the event horizon
- or crushed into free space according to the current development)
the geometry must be flat.

Finally, I would like to point out that the phase phase approach
considered in this paper requires one to think of space-time and
matter as unified, so I am not sure that it even makes any sense
within the present work to talk about a source of a gravity field
that is separated from the external field. The phase space
development keeps the source and the space-time around the source
united by replacing the singularity with the space-time-matter
energy per unit length calculated in equation 4.6.  According to
equation 4.6 the state of the matter in a ball of space-time-matter,
if any, is irrelevant, since the energy per unit length of radius is
constant.

Returning now to equation 4.3, since $r=sc=s$ if $c=1$, we can write
$m=ur=us$ to get,

\begin{equation}
\hat{\mathbf{P}}=\frac{\left[(\sqrt{1-u^2}-u^2)+i(u\sqrt{1-u^2}+u)\right]}{\sqrt{1+u^2}}\gamma<1,v>
\end{equation}

In this form the real and imaginary part of $\hat{\mathbf{P}}$ have
a very interesting property, namely, if
\begin{equation}
u=\frac{\sqrt{2\sqrt{5}-2}}{2}\approx 0.78615,
\end{equation}
then the real part of $\hat{\mathbf{P}}$ is zero, and the imaginary
part takes its maximum value ($=1$).  It is unclear to me exactly
how gravity manifests itself in the phase space approach, but I
think it makes sense to argue that when the real part of
$\hat{\mathbf{P}}=0$, $\hat{\mathbf{P}}$ is entirely ``mass like",
which we could understand to be representative of the state of
space-time-matter for which the maximal gravity field occurs. In
this picture gravity is understood to be the propensity of
space-time-matter to become completely mass like.  The more
mass-like a region of space-time-matter is, then the stronger the
external gravity field. Thus, within the discussion of this paper, I
think the only reasonable interpretation of the existence of the
special value of $u$ given in equation 4.10 is that there is a
maximal gravity field at this value of $u$.

The unique value of $u$ in equation 4.10 is only valid for the
hatted variables:  The real part $\mathbf{P}$ has a zero at this
value of $u$, but the imaginary part of $\mathbf{P}$ has a maximum
at a larger value of $u\approx 0.87$.  Yet another argument why the
hatted variables are a more natural choice.

Of course, a solid conceptual challenge of the present development
is that the parameter $u$ is not very intuitive, whereas the similar
parameter, density, is quite intuitive and constantly interferes
with thinking about $u$.  For example, for a very large region of
space-time-matter, say most of the observable universe, $u$ might be
greater than 0.1, even though the density of matter is near 0.  In
the phase space approach a large region of space-time-matter and a
compact region of space-time-matter have an immediate difference in
the expression for $\hat{\mathbf{P}}$, namely the value of $s$.  So,
if $s$ is large then the region of space-time-matter is large, etc.
In any case, in the statement ``how far a region of
space-time-matter is from being completely space-time like" we have
to be very careful not to confuse this with density.

It is important to observe that the value of $u$ considered above,
substantially exceeds the value of $u$ for a typical neutron star
($\approx 0.1-0.2$). Thus, I think the maximal gravity field concept
can be used to explain all of the experimental evidence for enormous
gravity fields.  For example, the best evidence for super black
holes asserts that $\sim 3.7$ million solar masses reside at the
center of our galaxy, inside of a ball the size of the inner solar
system, approximately extending out to Jupiter ~\cite{rS02,aG03}.
This is certainly an unimaginable region of space, but is far from a
black hole: $u\approx 0.02$, so we are not even in the neutron star
range. For 3-4 million solar masses to form a black hole they would
need to be confined to a region with a radius of about 9 - 12
million kilometres. Thus, the latest evidence only directly supports
the black hole singularity idea if we invoke a theory (general
relativity) that says such a massive compact object must collapse to
a singularity.

Although the phase space approach discussed in this paper is not
intended to be a theory of gravity, the concept of a maximal gravity
field seems to force a certain structure on a region of
space-time-matter with $u=\frac{m}{r}$ equal to the maximal value.
Apparently, for such a region to be stable, $u$ needs to have the
maximal value for any internal value of $r$, i.e., for any ball of
space-time-matter interior to the region we need to have $u$ equal
to the maximal value for the whole region to be stable. Then the
density function with respect to the observer's rest frame at the
center, should be structured to keep $u$ constant for any internal
value of $r$.  If the region has radius $R$ and mass $M$, then it is
straight forward to show that the density function needs to be
$\rho(r)=\frac{3M}{4\pi Rr^2}$. With this density function for any
$r$ inside the region of space
$\frac{m}{r}=\frac{\rho(r)V(r)}{r}=\frac{M}{R}$, where $V(r)$ is the
volume of a ball of space of radius $r$.

Therefore, as the observer sights along a radial line from the
origin, the density of matter inside the region is observed to drop
as a function of the inverse square.  Of course, it is not clear
exactly what this $r$ means, since our experience with general
relativity is that the gravity field affects our observation of $r$.
However, it seems reasonable to suppose that at least
asymptotically, $r$ has some resemblance to ordinary radial
distance.

However that may be, the above density function suggests that a
region of space with the maximal $u$ need not have a hard surface.
This is important, because accretion disk theory suggests that the
hard surfaces of neutron stars and the event horizons of black holes
should be observationally distinguishable ~\cite{NGM97,gB98}.

In addition, it is hard not to wonder if there is some sort of
duality relationship between the parameter $u$ and the inverse
square law for gravity (in the Newtonian limit).  Recall that we are
imagining that gravity is the propensity of a region of
space-time-matter to be mass-like, so if the density of matter is
reduced as the inverse square of $r$, this might explain why the
gravity field decreases as the inverse square of $r$.

Another point of interest is that the energy in free space is not
``mass like", because if it were, then free space would have a
gravity field so large that it would be opaque.  I think the phase
space approach offers a way-out of the problem posed by the
prediction of enormous amounts of background energy (both the
present approach and quantum field theory):  space-time energy
simply does not generate a gravity field.  Only, mass like energy
generates gravity fields. Specifically, only the mass part
(imaginary part) of $\hat{\mathbf{P}}$ generates a gravity field.

\subsection{Violation of Conservation of Energy}

As already mentioned, one of the most startling predictions of the
phase space development given here is that in extremely high gravity
fields conservation of energy apparently does not hold. What is
conserved for all values of $u$ and $v$ is the energy to length
ratio of any region of space-time-matter, $\frac{c^4}{G}$. Of
course, if $u$ is much less than 0.1 or so, the general phase space
reduces to the special case of Minkowski phase space, where
conservation of energy is correct to a high degree of approximation
(I think deviations are not experimentally verifiable).  The energy
equation, 4.5 reduces to $\frac{c^4}{G}$ for $u<<1$.  For example,
for a 1 kilogram mass with assigned proper time, $s$, such that
$sc=1$ metre, the energy calculated using equation 4.5 differs from
$\frac{c^4}{G}$ by about $3\times 10^{-53}$ percent. Since the
extreme high gravity environment is so unfamiliar experimentally, it
is hard to know what this conclusion means, but apparently we
experience conservation of energy in the low gravity environment,
because the space-time energy is so enormous and so far
undetectable.  Hence, we only detect masslike energy, unmixed with
space-time energy, so energy appears to be conserved, i.e., $E=m$
for all Lorentz observers.

\subsection{Symplectic Relativity Revised}

In the usual theory of special relativistic, Hamiltonian Mechanics
(non-quantum) the Hamiltonian is constant, i.e., $H=m$, where $m$ is
the rest mass of the particle. In the current development,
$H=\sqrt{s^2+m^2}$, so we can continue the development with a time
dependent Hamiltonian, even within special relativity. Actually, we
can treat the rest mass as a variable too, deriving a two parameter
special relativistic symplectic mechanics. Apparently, a better
choice would be to use the hatted Hamiltonian,
$\hat{H}=\frac{c^4}{G}$, but I am not sure how hatted symplectic
mechanics works.  A complex Hamiltonian, such as,
$\hat{H}=\frac{s+im}{\sqrt{s^2+m^2}}$ also seems possible.
Obviously, if the special relativistic phase space approach is
viable, it would be a good plan to work-out some of these ideas
before exploring the impact on quantum mechanics.  Evidently, we
need to use hatted energy and momentum operators and I don't know
what these would be. However, the last section of this paper may
help.

\section{Some Further Possible Developments of the Special Relativistic Phase Space Approach}

In this section we will consider two possible further developments
of the special relativistic phase space, namely adding more
parameters and deriving equations of motions from the Lagrangian
given in equation 3.3.  In both these discussions, $v$ remains
constant. Another possibility that I haven't discussed in this paper
is extending the phase space to general relativity, i.e., a
simplectic space in which both proper mass and proper time are
included as parameters. I just don't understand the present phase
space approach well enough at this time, to know how to extend it to
general relativity.
\subsection{More Parameters}

Having taken the plunge of forming a two parameter complex phase
space, merging space-time with matter, it is reasonable to wonder if
that is all we can do. For example, an obvious and appealing choice
is to add the charge to mass ratio, $w=\frac{q}{m}$, to our list of
parameters, $v$ and $u$ (adding the current would be another
possibility). There is even a convenient cluster of constants with
dimension of charge to mass, namely, $\sqrt{\epsilon_o G}$, where
$\epsilon _o$ is the permittivity of free space. At first glance
this cluster of constants will not do, because in metric units, the
value of $\sqrt{\epsilon_o G}\approx10^{-11}$ , whereas the the
charge to mass ratio of an electron is $1.76\times
10^{11}\mbox{coul/kg}$. Of course, the charge to mass ratio of many
particles (and large bodies) is zero, so apparently
$\sqrt{\epsilon_o G}$ cannot serve as a boundary value (lower or
upper) in the sense of $c$ and $\frac{c^2}{G}$. However, on further
reflection, in the phase space development, $w=\frac{q}{m}$ or the
current, $\frac{q}{s}$ are not really appropriate parameters, since
we have ignored the whole point of the phase space approach, namely,
that space-time and matter have been unified.  Thus, the parameter
that we want in the phase space approach of this paper is a combined
parameter, such as, $w=\frac{q}{\sqrt{s^2+m^2}}$ or possibly
$w=\frac{q}{s+im}$. Maybe an even better choice is a hatted
parameter, $\hat{w}=\frac{qG}{rc^2}$, where $r$ is the relativistic
Compton radius for the phase space.  We have calculated $r$ below.

For any of these parameters, $w$ (or $\hat{w}$) can be reconciled
with the ``boundary" cluster of constants if we include the solid
angle factor in the cluster to get, $\sqrt{4\pi\epsilon_o G}$. To
make the reconciliation we first need to recalculate the standard
relativistic expression for the Compton radius of an electron,
\begin{equation}
r=\frac{e^2}{4\pi\epsilon_o mc^2}=2.817940325\times 10^{-15}.
\end{equation}
using the phase space approach of this paper.  The factor of $mc^2$
in the denominator needs to be replaced with the phase space
expression for the total energy per unit length of a region of space
filled by the electron, given in equation 4.6. So we want to replace
$mc^2$ with $\frac{rc^4}{G}$. This gives,

\begin{equation}
r=\frac{e^2G}{4\pi\epsilon_o rc^4}
\end{equation}

Solving for $r$, we get,
\begin{equation}
r=\frac{e}{c^2}\sqrt{\frac{G}{4\pi\epsilon_o}}\approx
0.13767776994\times 10^{-35}\mbox{metres}
\end{equation}

This seems like an interesting result: the phase space approach
appears to predict that the electron Compton radius is only about
9\% of the Planck Length. Of course, it would be quite significant
if $r$ worked-out to exactly the Planck length.  I have searched for
a simple way to adjust the above calculation, but can't find a
suitable method. Obviously, we could start by normalizing both
lengths over the same angle (the Compton radius is averaged over a
solid angle, whereas the Planck length is averaged over a polar
angle), but doing this doesn't make the two lengths equal. Evidently
charge quantization is independent of spacial-energy quantization
measured by Planck's constant.  Why this is so when we have the
energy equation, $E=h\nu$, clearly also an electromagnetic equation,
is baffling. For the record, the ratio of the Planck Length $L$ to
the phase space Compton radius is
$\frac{L}{r_c}=\frac{\sqrt{2h\epsilon_o c}}{e}\approx
11.4675833875$.

In any case, the extended value of $r$ is presumably a great success
of the phase space approach, since quantum field theory works best
if $r$ is just assumed to be zero. There is a problem with this
calculation that perhaps makes it not particularly useful, namely,
the vast majority of the energy in the expression $E=\frac{rc^4}{G}$
is not accessible by any known means (at least, short of applying
immense gravity fields), so it will not really contribute in a
practical experiment. So perhaps equation 5.1 will remain the
practical Compton radius. This is also likely, simply because the
phase space version of the Compton radius does not depend on the
mass of the particle. It is perhaps worth mentioning that the
utility of the relativistic Compton radius calculated in equation
5.1 must mostly be a coincidence, since the energy factor used in
the denominator, $E=mc^2$ is not accessible except for the highest
energy experiments, e.g., an electron-positron annihilation.

Continuing with the calculation of the relativistic phase space
``charge to mass ratio" of an electron:  using $r$ from equation 5.3
in $w=\frac{q}{\sqrt{s^2+m^2}}$, and inserting the appropriate
constants ($q=e$), $w$ works out to $w=.8613270258\times 10^{-10}$.
Comparing, this is exactly equal to, $\sqrt{4\pi\epsilon_oG}$, at
least to ten decimal places. The electron ``charge to mass" ratio is
actually smaller than $\sqrt{4\pi\epsilon_oG}$, but according to
Maple the difference is of order $10^{-80}$.

Alternatively, if we define $w=\frac{q(s-im)}{s^2+m^2}$ then the
real part works out to $.8613270258\times 10^{-10}$ and the
imaginary part works out to $-.4223547307\times 10^{-31}$.  Then
$|w|=.8613270258\times 10^{-10}$.  This time $|w|$ differs from
$\sqrt{4\pi\epsilon_oG}$ in approximately the fortieth decimal
place.  The hatted version works-out to exactly,
$\hat{w}=\sqrt{4\pi\epsilon_o G}$, which isn't surprising.

Unfortunately, it does not look like $w$ (any version) is going to
be a very useful parameter, because the relativistic phase space
Compton radius does not depend on the mass. So
$\hat{w}=\sqrt{4\pi\epsilon_o G}$ for any particle with non-zero
charge. Evidently, in the space-time-matter-charge configuration
space, in geometrized units ($4\pi\epsilon_o=1$), the ``charge to
mass ratio" of any particle with non-zero charge is 1.  This is
apparently similar to the fact that in the usual theory of special
relativity, the four velocity has magnitude -1 for any particle.
Probably the complex version of $w$ will be the most useful, because
we can study the components.

More generally, the difference in magnitude between the standard
value of the charge to mass ratio of an electron, $1.76\times
10^{11}\mbox{coul/kg}$, and cluster of constants,
$\sqrt{4\pi\epsilon_oG}$, seems to refute the idea that clusters of
constants serve as boundary values for relativistic parameters. In
particular, without the possible reconciliation available with the
phase space approach, even the idea of $u$ and its boundary of
$\frac{c^2}{G}$ would be suspect, since we would have found a simple
counterexample in $w$. So the reconciliation explained above is of
theoretical interest, since it saves the day for the idea that
certain relativistic parameters are bounded by universal constants
(or clusters of constants) that I have assumed have the same values
for all observers.  In any case, we can continue with a sketch of
the phase space development with more parameters.

The next important step would be to decide on a number system with
which to express the additional parameters.  The quaternions seem to
be the best bet, but I don't know what to do with the fourth
generator (only three generators seem to be needed: one for each of
$v$, $u$, and $w$).  The fourth generator needs to be a physical
quantity commonly parameterized by all of the proper time, proper
mass and the charge. Also, the quaternions have anti-commutative
multiplication between imaginary generators, and I don't know what
that means (why would $ui\times wj=uwk=-wj\times ui$?).  Perhaps
this is just what we need, since the metric we have been using for
section 4 and 5 of this paper is actually a symplectic two form,
which also anticommutes. The numbers that we use need to have the
complex numbers as a subfield, so for example, the Dirac or Pauli
matrices are out. In any case, we can construct the velocity vector
in an extended phase space in $Q^2$, (Q for quaternions):
\begin{equation}
\mathbf{U}=(\sqrt{1-u^2-w^2-z^2}+iu+jw+kz)\gamma<1,v>
\end{equation}

In this equation $i, j,$ and $k$ are the generators of the
quaternions, and $z$ is the unknown fourth parameter.  This
expression is going to have to be interpreted very carefully,
because charge is invariant in special relativity.  In any case, we
can conjecture that the momentum form is given by:

\begin{equation}
\hat{\mathbf{P}}=<y^0,y^1>=(\sqrt{1-u^2-w^2-z^2}+iu+jw+kz)\gamma<1,v>\frac{s+im+jq+kn}{\sqrt{s^2+m^2+q^2+n^2}}
\end{equation}

In this equation $n$ is the proper parameter that is associated with
$z$. Then it is tempting to conjecture that the revised energy
equation will just work-out to $E^2=s^2+m^2+q^2$ (assuming $n=0$),
using units in which $c$, $G$ and $4\pi\epsilon_o$ are all 1. The
justification for this conjecture is that whatever the actual
formulation is, the resulting tangent vector (form) will have
magnitude of -1. Thus, there will be some extended momentum vector,
and $E^2$ will still be equal to $-\mathbf{P}\cdot\mathbf{P}$.
Inserting the appropriate constants,
\begin{equation}
E^2=\frac{s^2c^10}{G^2}+m^2c^4+\frac{q^2c^4}{4\pi\epsilon_o G}
\end{equation}

Presumably this expression will need to be re-scaled in some way to
give equation 4.6.  But for now we can see that the contribution to
the total energy from the charge term in equation 5.4 is
$E=\frac{qc^2}{\sqrt{4\pi\epsilon_o G}}$. For $q=1.6\times
10^{-19}\mbox{coul}$, $E\approx 1.3\times 10^{3}\mbox{joules}$,
which is enormous (approximately 16 orders of magnitude larger than
the mass energy of an electron). However, the phase space approach
gives no clues as to how this energy can be accessed. We can
interpret this result as simply the stored energy due to the charge,
in the same way that the phase space approach developed in this
paper predicts that there is a huge amount of energy stored in free
space. So far we do not know how to access the energy stored in free
space, either.  It is straight forward to show that the ratio
between the charge energy and the space-time energy per unit length
is just the phase space Compton radius that we found before
(equation 5.3).

Notice that we will most likely need to use a parameter with unit
modulus, so equation 4.6 still holds. We then have space-time
energy, charge energy, and mass energy all intermixed in any region
of space-time-matter-charge.

We haven't discovered how to include charge in such a way that the
charge remains invariant (with respect to changes in $v$) in the
appropriate special cases, so the suggestions here can't be correct,
but gives some idea of how a further development in this direction
might go.

\subsection{Derivation of Equations of Motion}
As previously observed the phase space Lorentz transformation
(equation 3.19) can be derived simply using a complex conjugated
metric, but one reason to use the Lagrangian in equation 3.3 is so
we can apply an action principle to derive a set of equations which
we can then solve to find a general expression for $\mathbf{P}$.
First recall equation 3.3:

\begin{equation} L^2=\delta_{kl}^{01}\delta_{np}^{01}
(y^k_{, s}y^l_{, m})^{\dagger}y^n_{, s}y^p_{, m}
+\delta_{kl}^{20}\delta_{np}^{20} (y^k_{, s}y^l_{,
m})^{\dagger}y^n_{, s}y^p_{, m}- \delta_{kl}^{12}\delta_{np}^{12}
(y^k_{, s}y^l_{, m})^{\dagger}y^n_{, s}y^p_{, m}
 \end{equation}

 In this expression the $y^k$'s are understood to be the components of the
 generalized phase space (simplectic) momentum vector $\mathbf{P}$ or $\hat{\mathbf{P}}$.  I have already given one
 special case for $\hat{\mathbf{P}}$ in definition 4.2.  Then the $y^k_s$ and $-iy^k_m$
 partials are tangent vectors.  The factor of $-i$ in front of the $m$ partial
 arises because $m$ is the imaginary part of the complex parameter,
 $\sigma=s+im$.
 To keep the equations as simple as
 possible, we will continue in accordance with equation 3.19, and assume
 that $y^2=1$.
 For the following derivation I am assuming that the particle
 follows a stationary path defined by the Euler-Lagrange equations
 for $L^2$.  Recall ~\cite{GPS02} that these equations are as follows:
 \begin{equation}
\frac{\partial L^2}{\partial y^k}-\frac{D}{Dm}(\frac{\partial
L^2}{\partial y^k_m})-\frac{D}{Ds}(\frac{\partial L^2}{\partial
y^k_s})=0
\end{equation}

In these equations $\frac{D}{Dm}$ and $\frac{D}{Ds}$ are ``total"
derivatives, i.e., the chain rule should be applied when
differentiating $\frac{\partial L^2}{\partial y^k_m}$ and
$\frac{\partial L^2}{\partial y^k_s}$.  We have used the special
notation, because for example, the expression $\frac{\partial
L^2}{\partial y^k_m}$ doesn't have explicit dependence on $m$, so
the ordinary partial of this expression with respect to $m$ would be
zero.  This is not what we want.  Also, in the second term on the
left hand side of equation 5.6, factors of $i$ that arise from
differentiating with respect to $m$, have canceled.  For simplicity,
I am assuming that $k=0,1$.  In any case, substitution of equation
5.5 into equation 5.6 and simplifications yields the system of
partial equations that govern the motion of the particle in the
special relativistic phase space discussed in this paper.
\begin{eqnarray}
(y^1_s)^2y^0_{mm}-y^0_sy^1_sy^1_{mm}-2y^1_my^1_sy^0_{sm}+(y^0_my^1_s+y^1_my^0_s)y^1_{sm}+\\
(y^1_m)^2y^0_{ss}-y^0_my^1_my^1_{ss}=0\nonumber
\end{eqnarray}
\begin{eqnarray}
(y^0_s)^2y^1_{mm}-y^1_sy^0_sy^0_{mm}-2y^0_my^0_sy^1_{sm}+(y^1_my^0_s+y^0_my^1_s)y^0_{sm}+\\
(y^0_m)^2y^1_{ss}-y^1_my^0_my^0_{ss}=0\nonumber
\end{eqnarray}
Since equations 5.7 and 5.8 both involve second order partials in
every term, the momentum vector defined in 4.2 obviously satisfies
both equations.

The system of equations in 5.7 and 5.8 has many solutions.  For
example, $<f(s+im),f(s-im)>$ satisfies both equations for any twice
differentiable $f$ (differentiable in the sense of multi-dimensional
calculus-$f(s-im)$ is obviously not analytic, unless $f$ is
constant). We would like a solution that linearizes to the momentum
vector defined in 4.2. One possibility is,
\begin{equation}
<y^0,y^1>=<\gamma(\exp((\sqrt{1-u^2}+iu)(\frac{s+im}{\sqrt{1+u^2}}))-1),
v\gamma(\exp((\sqrt{1-u^2}+iu)(\frac{s+im}{\sqrt{1+u^2}}))-1>
\end{equation}

It can be readily verified that any vector of the form,
\begin{equation}
<y^0,y^1>=<A\exp(a(s+im)),B\exp(b(s+im))>
\end{equation}

satisfies 5.7 and 5.8, where $A,B,a$ and $b$ are arbitrary (complex)
constants. Hence, the solution in equation 5.9.

Frankly, I do not understand what the solution in equation 5.9
means. Certainly, $v$ is constant, so this equation is not about
actual motion, just configuration space motion.  If
$\hat{\mathbf{y}}=<y^0,y^1>$, then certainly both the tangent
vectors, $\hat{\mathbf{y}}_s$ and $\hat{\mathbf{y}}_m$ have
magnitudes of $-1$, using the complex conjugated Lorentz metric
(explained section 4). It is entirely unclear to me what kind of
boundary conditions to impose on the system in 5.9.

Since the constants, $a$ and $b$ in equation 5.10 do not need to be
real, we must have wave solutions as well. In fact, the solution in
5.9 is already a wave with an increasing amplitude.

\section{Conclusion}
In this paper we have developed a complex phase space within the
context of special relativity. Minkowski space (space-time) is
changed to a space-time-matter configuration space in $C^2$ or
($C^4$), by including a complex phase factor in the Lorentz metric
and allowing the proper mass to be treated as a parameter.
Unfortunately, the limiting case for which $u\approx 1$ is the
domain of enormous gravity fields, so full interpretation of the
results will probably require generalizations of the techniques
discussed in this paper to general relativity. Nevertheless, the
phase space approach seems to provide an alternative to
gravitational singularities: matter is crushed into free space as
$u$ increases past $u\approx .79$ and there is a stable maximum
gravity field with a soft surface.

According to the development in this paper, the special relativistic
phase space expression for the energy is $E^2=s^2+m^2$. Among other
things, this equation says that ordinary space-time contains an
unfathomable amount of energy. Apparently, the quantity that is
invariant in the phase space is the energy to length ratio,
$E=\frac{c^4}{G}$.  I have discussed a number of other consequences
as well, such as the baffling similarity (but not equality) of the
phase space Compton radius and the Planck length.

Finally, I have considered some further developments of the
approach. It certainly seems likely that more parameters can be
added, that an interesting two parameter Hamiltonian mechanics can
be constructed (using either energy or energy per unit length),
which should have consequences for quantum mechanics, and that the
general equations of ``motion" in space-time-matter have many
possible solutions.

\bibliographystyle{amsplain}
\bibliography{extrel}
\end{document}